\documentstyle[12pt,openbib]{article}
\newcommand{\G}{\Gamma}
\newcommand{\g}{\gamma}
\newcommand{\de}{\delta}
\newcommand{\be}{\beta}
\newcommand{\f}{\frac}
\newcommand{\ka}{\kappa}

\newcommand{\p}{\prime}
\newcommand{\r}{\rho}
\newcommand{\om}{\omega}
\newcommand{\ep}{\epsilon}
\newcommand{\al}{\alpha}
\begin{document}
\begin{center}
{\Large{Limiting temperature of hadrons using  states predicted from
$\ka $-deformed Poincar\'e algebra}}
\end{center}
\vskip .5cm
\begin{center}
Jishnu Dey  $^{2,3}$, Siddhartha Bhowmik $^{1}$, Kanad Ray $^{1,2,3}$ and 
Subharthi Ray $^{1,2,3}$
\end{center}

\begin{center}
{\it Abstract}
\end{center}
The experimental hadronic density of states dN/dm, assumed to be a sum of
normalized Breit- Wigner distributions and plotted as a function of the
hadron mass $m$, fails to show a Hagedorn like growth beyond 2 GeV,
probably due to a lack of data. Experimental  hadronic states are
fitted using $\ka$ -deformed Poincar\'e algebra and the fit is used to
extrapolate for including states not detected. For the theoretical density of
states the plot is a straight line in the log scale even beyond 2 $GeV$ with
a limiting temperature of 400 $MeV$.

\vskip 1cm 
\noindent
(1) Dept. of Physics, Presidency College, Calcutta 700 073, India \\
(2)Azad PhysicsCentre, Dept. of Physics, Maulana Azad College, Calcutta 700 013, India\\ 
3) {\it{ Work supported in part by DST grant no. SP/S2/K18/96, Govt.
of India}}
\newpage

Deformed Poincar\'e algebra, (dPa in short), keeps the three dimensional rotation and the
translation subgroups undeformed while the algebra of Lorentz boosts is
modified, both for bosons and fermions. The relevant q-deformation parameter
is called $\ka $ in this case and when this goes to infinity we recover the
undeformed algebra.  The $\ka $-deformed Dirac equation has
recently been found \cite{no}.  Extensive applications of dPa have
been carried out to see what would be its impact on the standard theories
governed by the ordinary quantum special relativity. The following problems
have been studied.

a) The definition of mass with different non-relativistic limits
\cite{bac1}, 

b) the non-additivity of masses and its relation to the
interesting dark matter puzzle \cite{bac2}, 

c) the classical electrodynamics problem of finding the acceleration
of charged particle in a one-dimensional homogenous electric field
\cite{bac3},

d) gauging the deformed Dirac equation, applying it to the quantum
relativistic hydrogen atom and solving the Dirac-Coulomb problem \cite{bi},

e) calculating the Landau deformed levels \cite{rr},

f) explanation of the flattening of the experimental hadron spectrum 
\cite{de}, \cite{ddft},

g) application of the new mass-energy relation of $\ka $-deformed
algebra to the model of Nambu and Jona-Lasinio,
{\it now with a natural cut-off $1/\ep $ provided by the theory}
\cite{del}.  

h) Quite recently it has been suggested \cite{cp} that matter and radiation
can be created in the confined vacuum of a quantum field whose spacetime
symmetries are governed by  Poincar\'e algebra. It is claimed that the
creation rate goes to zero when the deformation disappears. We shall have
occasion to come back to a further discussion of this very interesting paper.

i) The flattening of hadron spectrum, explained by the deformed algebra in
the case (f), seems to lead to interesting smooth phase transitions at finite
T \cite{cdd}.

From one of these studies, namely the case (d), it turns out that for
negligible deformation, the normal Dirac equation is recovered.  Expansion in
the deformation parameter gives the result that the first order effect
vanishes identically \cite{bi}. This means clearly that there is no change
in the energy spectrum in the first order of perturbation theory.  This does
not happen for the deformed Landau levels \cite{rr} , which are expected to
shift already in first order perturbation theory. As we can see, people are
getting interested to see how a determined theory or equation behaves under a
new symmetry structure generated by a group deformation. 

In the present paper we use the formalism described above to fit and
extrapolate the observed baryons. Some of the  mesons were already fitted
\cite{ddft}, we fit the rest, viz. the $K$ and the $K^*$ and the $\eta$ and
the $\eta^{\p}$.  We use
\begin{equation}
M(n, L, S, J) = \f{2}{\ep } sinh^{-1} \left[  { \left( \f{\ep}{2}\right) ^2
\left(\f{L}{\al ^\p  } + \f {n} {\be ^\p  } + \f {S} {\g ^\p  } + \f {J} {\de
^\p  }\right) + sinh^2 \left( \f{m\ep}{2}\right) }\right]^{1/2}.
\label{eq:1}
\end{equation}
where $L, S$ and $J$ stand for orbital, spin and total angular momentum and n
is the quantum number for radial excitation.

The value of $\ep$ is fixed once for all at $0.915\; GeV^{-2}$. For the
$\pi, \r \; \rm and \; \om$ we use $m \; = \; 0.138\; GeV$. This implies that
the $\r - \om$ are spin-excitations of the pion. In the same way we use $m \;
= \; 0.494\; GeV$ for $K$ and $K^*$. The spin parameters $\g ^\p \; = \; 2.35
\; GeV^{-2}$ and $\de ^\p \;=\;5.5\;GeV^{-2}$ are also unaltered. The
parameters $\al ^\p \; = \; 0.7 \; GeV^{-2}$ and $\be ^\p \; = \; 0.5292 \;
GeV^{-2}$ for $\pi$, $\r$ and  $\om$ are changed to $0.679$ and $0.44$ for
the strange mesons. For the $\eta$ and $\eta^{\p}$, $m$ is 0.547 and 0.958
GeV and $\al ^\p \; = \; 0.99 \; GeV^{-2}$ and $\be ^\p \; = \; 0.67 \;
GeV^{-2}$.

For baryons the $\de ^\p$, $\al ^\p$ and $\be ^\p$ are given in their
respective tables.

A series of papers, \cite{fr}, \cite{cd} and \cite{dc}, deal with the
densities of observed mesons and baryon states and their possible
relationship with hadron-scale string theories. The frustration involved in
this kind of work stems from the fact that the experimental states are known
only upto $\approx 2.5 \;GeV$ and even in this region probably many states
are not experimentally identified. Thus the total density of hadrons in Fig.
1 plotted in log scale fails to grow linearly beyond 2 GeV, and this is
`likely to be a reflection of current experimental limitations'
\cite{cd}. Since we are able to predict meson and baryon states, we check
this result. Indeed Fig. 1 with the extrapolated states goes like a straight
line with a slope of $T_H \;=\; 400$ $MeV$, slightly larger than the values
of 250 \cite{cd} or 300 \cite{dc} $MeV$, but quite in line with the
expectations of Cudell and Dienes. Note that the value $T_H \;\sim \;160\;
MeV$ for the Hagedorn temperature is too low to agree with the central charge
of the effective QCD string \cite{dc}. 

To plot Fig. 1 we use :
\begin{equation}
\f{dN}{dm}= \f{1}{2 \pi} \sum _{i}W_i \f{\G_i}{(m-M_i)^2 +
\f{\G_i^2}{4}}
\label{eq:2}
\end{equation} where the masses $m$ and widths $\G$ are taken from \cite{pdg}
for the experimental curve (with ++). For the theoretical curve (with dots).
we use the masses from eqn. (\ref{eq:1}) with widths $8.5\; MeV$ below 1 GeV
and $55\; MeV$ above. This choice makes the dots relatively smooth.
The important point is that from 2 to 3 GeV, the theoretical curve smoothly
fits onto  $\sim exp (m/T_H)$ with $T_H \,= \,400 \;
MeV$

Eqn. (\ref{eq:1}) fits the experimental states rather well. The $\pi,
\eta, \eta^\p$ and $K$ are fitted and therefore left out of table 1. Note the
good fit to the $\r, \om, K^*, h$(first and second), radial excitations of
$\pi$ at $1.3 $, $K$ at 1.46, $(\r,\; \om)$ at $(1.7,\,1.6)$, - even $\r _5$
at $2.35$ and $K_4^*$ at 2.045 (all in $GeV$).

For the baryons the ground states, which are fitted, are also put in the
table to enable the reader to identify the sets easily. For nucleon states
the Roper at $1.44$ and its higher radial excitations are well fitted, but
there is the well-known problem of fitting the second S11 state, while the
third S11 is well fitted.  The other angular excitations are also reasonably
well fitted and we are anticipating new experimental data to come from CEBAF
(Jefferson centre). For the strange baryons the fit is similar in quality.

~~We next turn to thermodynamics of the hadron gas. In \cite{cdd} it was
suggested that there is a smooth phase transition in energy density in the
extrapolated hadron spectrum using deformed Poincar\'e algebra. However it is
now clear to us that thermodynamic quantities are ill-defined and the sum
over particle states in them do not converge beyond $T_H$.
 It is
also clear  that dPa applies only to the internal structure
of the hadrons. However, below $T_H$ we can still calculate the free energy F
(and the energy E), of the hadron gas :
\begin{equation}
F(T)\,\,=\,\frac{T}{2\pi\,^{2}}\int_{0}^\infty\,\,k^{2}
\sum_{nLSJ}\,\,g_{nLSJ}\,\,ln\,[\,1\,-\,e\,^{-E_{k}/T}\,]\,dk
\label{eq:3}
\end{equation}
for Bose gas and 
\begin{equation}
F(T)\,\,=\,\,-\frac{T}{2\pi\,^{2}}\int_{0}^\infty\,\,k^{2}
\sum_{nLSJ}\,\,g_{nLSJ}\,\,ln\,[\,1\,+\,e\,^{-E_{k}/T}\,]\,dk
\label{eq:4}
\end{equation} 
for Fermi gas with
\begin{equation}
E_k =\,\sqrt {k^2 + M(n, L, S, J)^2}
\label{eq:5}.
\end{equation} 

Hence its entropy $S \equiv (E\,-\,F)/T$ is known. We plot $ (E - F)/ E$ in
fig. 2. This quantity obviously starts from zero. It fast approaches the
value 1.2 for mesons, almost independent of the temperature, tantalizingly
close to the ratio $4/3$ as in the case of massless quarks and gluons. There
is a little dip in the curve which (fig.2) we do not understand at present
and do not wish to comment on. For baryons the ratio is somewhat less, close
to 1.1, but almost independent of T in the range displayed. Bearing in mind
that at least the strange quark {\it is} massive, it may be possible that
somewhere below the $T_H$ a realistic quark -gluon description sets in rather
smoothly. We find it interesting that the hadron gas, with the high
occupation probability of massive resonances, still has a $ (E - F)/ E$ ratio
which is close to that of nearly massless particles.

In summary we find the string theorist's expectation that the Hagedorn $T_H$
is almost double the conventional value $\sim 160\; MeV$ is borne out for hadronic states generated by $\ka$ -deformed Poincar\'e algebra. 
This work was supported in part by a grant from the Department of Science and
Technology, Govt. of India, two of the authors (Rays) hold appointment under
this grant.   

\begin{table}
\begin{center}
\caption {Meson masses from our model compared to experiment}
\vskip .2cm
\begin{tabular}{|c|c|c|c|c|c|}
\hline
 Meson state & Ours &  Expt &Meson state & Ours & Expt \\
\hline
$\r$, $\omega$ & .775  & .77, .782 & $\r$, $\omega$ & 1.692 & 1.7, 1.6 \\
$\pi$2 & 1.641  & 1.67 & b1 & 1.213 & 1.235 \\
$\r$3, $\omega$3 & 1.764  & 1.69, 1.67 & a4, f4 & 2.031 &2.04, 2.05 \\
$\r$5 & 2.243  & 2.35 & a1, f1 & 1.347 & 1.26, 1.285 \\
a2, f2 & 1.398  & 1.32, 1.27 & $\r$3 & 2.071 & 2.25 \\
f2 & 1.815  & 1.81 & $\r$, $\omega$ & 1.472 & 1.45, 1.42 \\
$\r$ & 1.865  & 2.15 & $\pi$ & 1.303 & 1.3 \\
$\pi$ & 1.754  & 1.8 & $\pi$2 & 1.982 & 2.1 \\
f2 & 2.11  & 2.15 & f0 & 2.059 & 2.2 \\
f2 & 2.339  & 2.3 & f4 & 2.276 & 2.3 \\
 a6, f6 & 2.42  & 2.45, 2.51 & $\eta$2 & 1.939 & 1.87 \\
h1 & 1.167  & 1.17 & $\eta$  & 1.461 & 1.44 \\
$\eta$ & 1.269  & 1.295 & h1 & 1.38 & 1.38 \\
$\eta$ & 1.652  & 1.76& K1 & 1.3012 & 1.27 \\ 
K* & .899  & .892& K* & 1.755 & 1.68 \\ 
K1* & 1.423  & 1.4 & K3*  & 1.822 & 1.78 \\
K2 & 1.707  & 1.77& K5* & 2.291 & 2.38 \\ 
K4* & 2.083  & 2.045 & K2* & 1.927 & 1.98 \\
K2* & 1.471  & 1.43& K & 2.023 & 2.1 \\ 
K* & 1.617  & 1.68 & K & 1.929 & 1.83 \\
K & 1.474  & 1.46 & - & - & - \\
\hline
\end{tabular}
\end{center}
\end{table}
\begin{table}
\begin{center}
\caption {Different value of the parameters for the baryons}
\vskip .2cm
\begin{tabular}{c|c|c|c|c|c}
\hline
Baryon Name & $\de^\p$ & $\al ^\p$ & $\be ^\p$ & $\g^\p$ & m \\
\hline
Nucleon & -5.5 & .58 & .685 & 2.6 & .889\\
Delta & -7.2 & .48 & .685 & 2.6 & 1.12\\
Lambda & -5.5 & .7 & .8 & 2.6 & 1.077\\
Sigma & -5.5 & .7 & .8 & 2.6 & 1.153\\
Cascade & 1.15 & .7 & .8 & 2.6 & 1.102\\
\hline
\end{tabular}
\end{center}
\end{table}

\begin{table}
\begin{center}
\caption {Masses of baryons from our model  compared to experiment}
\vskip .2cm
\begin{tabular}{c|c|c|c|c|c|c|c|c}
\hline
 Nucleon state & Ours &  Expt & Nucleon state & Ours & Expt & Nucleon state & Ours & Expt \\
\hline
P11 & 0.939  & 0.939 & P11 & 1.765 & 1.71 &P11 & 2.01  & 2.1  \\
P11 &  1.441 & 1.44& P13 & 1.829  & 1.72& G17 & 2.046 & 2.19  \\
D13 & 1.463  & 1.52&P13 & 2.061 & 1.9 & D15 & 2.26  & 2.22  \\
S11 & 1.508 & 1.535&F17 & 2.269  & 1.99 &  H19 & 2.248 & 2.22  \\
S11 & 1.814 & 1.65 &F15 & 2.034 & 2.0& G19 & 2.435  & 2.25 \\
F15 & 1.796 & 1.68 &D13 & 2.022  & 2.08&I1,11 & 2.417 & 2.6 \\
D13 & 1.781  & 1.7 &S11 & 2.049 & 2.09&K1,13 & 2.563  & 2.7\\
\hline
 Delta state & Ours &  Expt & Delta state & Ours & Expt& Delta state & Ours & Expt \\
\hline
P33 & 1.232  & 1.232& P33 & 2.087 & 1.92& H39 & 2.51 & 2.3  \\
P33 &  1.621 & 1.6 &D35 & 1.976  & 1.93& D35 & 2.327 & 2.35\\
S31 & 1.775  & 1.62&D33 & 1.997 & 1.94 & F37 & 2.522  & 2.39  \\
D33 & 1.748 & 1.7&F37 & 2.048  & 1.95& G39 & 2.296 & 2.4   \\
S31 & 2.018 & 1.9 & F35 & 2.256 & 2.0& H3,11 & 2.497  & 2.42   \\
F35 & 2.068 & 1.905 & S31 & 2.214 & 2.15& I3,13 & 2.666 & 2.75   \\
P31 & 2.106  & 1.91 & G37 & 2.311  & 2.2& K3,15 & 2.812  & 2.95  \\ 
\hline
 Lambda state & Ours &  Expt & Lambda state & Ours & Expt & Lambda state & Ours & Expt\\
\hline
P01 & 1.116  & 1.116& S01 & 2.006  & 1.8 & F05 & 1.978 & 2.09 \\
S01 &  1.539 & 1.407& P01 & 1.766 & 1.81& G07 & 1.977  & 2.1 \\
D03 & 1.495  & 1.52 & F05 & 1.765 & 1.815 & D03 & 1.978 & 2.325 \\
P01 & 1.496 & 1.6& D05 & 2.005 & 1.83 & H09 & 2.153  & 2.35 \\
S01 & 1.799 & 1.67& P03 & 1.799  & 1.85& - & - & -    \\
D03 & 1.765 & 1.685 & F07 & 2.177 & 2.02& - & - & -   \\
\hline
 Sigma state & Ours &  Expt & Sigma state & Ours & Expt & Sigma state & Ours & Expt\\
\hline
P11 & 1.189  & 1.189&D15 & 2.034  & 1.77& S11 & 2.035 & 2.0  \\
D13 &  1.544 & 1.58&P11 & 1.803 & 1.77& F17 & 2.202 & 2.025\\
S11 & 1.586  & 1.62 &P13 & 1.835 & 1.84 & F15 & 2.008  & 2.07\\
P11 & 1.544 & 1.63 & P11 & 2.009 & 1.88 & P13 & 2.035 & 2.08   \\
D13 & 1.803 & 1.665& D13 & 2.009 & 1.9  & G17 & 2.008  & 2.1    \\
S11 & 1.835 & 1.73& F15 & 1.802  & 1.9  & - & - & -  \\ 
\hline
 Cascade state & Ours &  Expt & Cascade state & Ours & Expt
 & Cascade state & Ours &  Expt  \\
\hline
P11 & 1.315  & 1.315 & D13 & 1.837 & 1.823 & - & - & - \\
\hline
\end{tabular}
\end{center}
\end{table}

\vskip .1cm
\begin{figure}
\begin{center}
\input{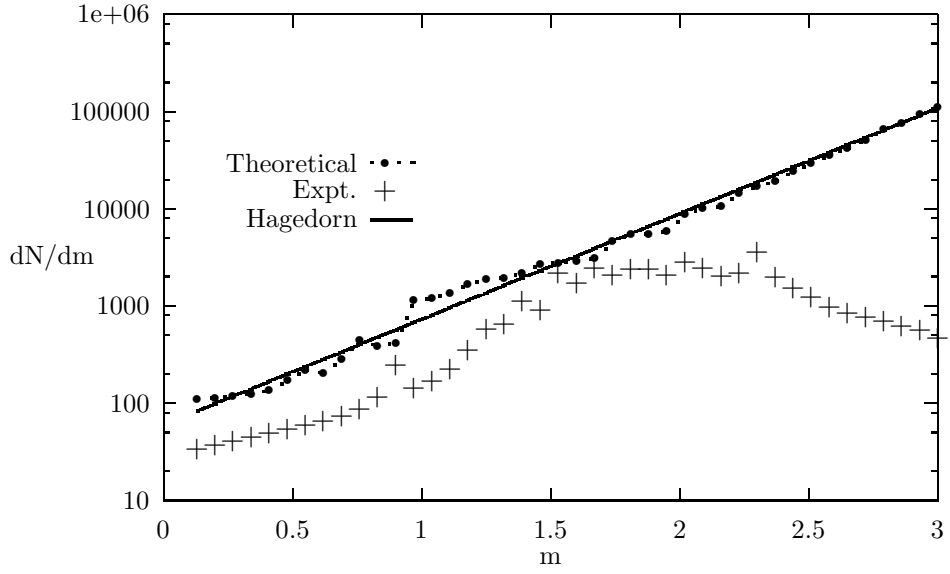}
\caption{Comparison of experimental (`plus'-s, ++) and theoretically
predicted density of states (dots ......) The straight line gives $\sim exp
(m/T_H)$.}
\end{center}
\end{figure}

\vskip .1cm
\begin{figure}
\begin{center}
\input{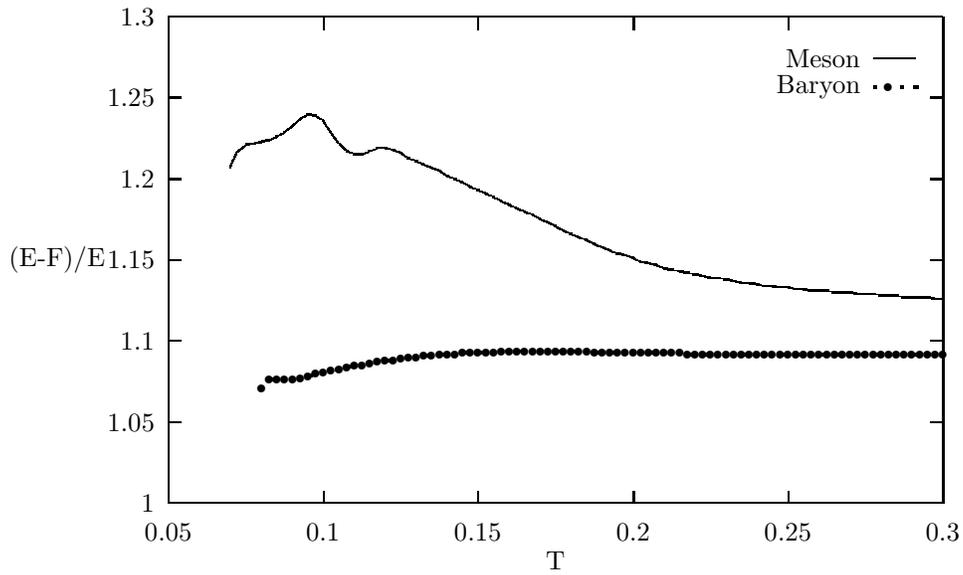}
\caption{The calculated values of\ \  $\f{E-F}{E}$\ \ for mesons and baryons. }
\end{center}
\end{figure}

\end{document}